# Capillary bridge formation and breakage: a test to characterize antiadhesive surfaces

TITLE RUNNING HEAD: Capillary bridge for antiadhesive surfaces.

*Laurianne Vagharchakian*[1], *Frédéric Restagno*[2], *Liliane Léger*[3]

Laboratoire de Physique des Solides, Univ Paris-Sud, CNRS, UMR 8502, 91405 Orsay Cedex, France.

[1] Author email address: vagharchakian@yahoo.com

[2] Author email address: restagno@lps.u-psud.fr

[3] Author email address: leger@lps.u-psud.fr

[3] To whom correspondence should be addressed. This work has been mainly performed at Laboratoire de physique des fluides organisés, FRE 2844 CNRS – Collège de France, 11 Place Marcelin Berthelot, 75005 Paris, France, before the closure of the lab when Pierre-Gilles de Gennes retired from Collège de France.

**ABSTRACT**

In order to characterize very weakly adhesive surfaces, we have developed a quantitative test inspired by the JKR adhesion test for soft adhesives, which relies on the formation and then the rupture of a capillary bridge between the surface to be tested and a liquid bath. Both the shape and the kinetics of breakage of the capillary bridge for various coatings put into contact with liquids of various viscosities and surface tensions have been studied. Several pull off regimes can be distinguished. For low pull off velocities, a quasi-static regime is observed, well described by capillary equations, and sensitive to the hysteresis of the contact angle of the fluid on the coating. Above a critical pull off velocity which depends on the fluid viscosity, a dynamic regime is observed, characterized by the formation of a flat pancake of fluid on the coating which recedes more slowly than the capillary bridge itself. After the breakage of the capillary bridge, a small drop can remain attached to the surface. The volume of this drop depends on the dynamical regime, and is strongly affected by very small differences between the coatings. The aptitude of this test in characterizing very weakly adhesive surfaces is exemplified by a comparison between three slightly different perfluorinated coatings.

**INTRODUCTION**

A number of practical situations such as anti-dirt or anti fouling coatings deal with extremely weakly adhesive surfaces. While the exact level of adhesion is often crucial for applications, such a very weak adhesion is difficult to characterize in a objective manner: essentially all adhesion tests now available for soft adhesives and moderately weak adhesion lead to immediate failure on these anti-adhesive coatings, providing no quantitative way of determining an adhesion energy. All these common adhesion tests, such as for example the now widely used Johnson, Kendall and Roberts (JKR) test[1] do oppose elastic deformation forces to adhesion forces. With very weakly adhesive surfaces, even soft elastomers have a too high elastic modulus to provide a balance between elastic and adhesive energies for detectable strains. In order to overcome this difficulty and gain a better insight into the molecular mechanisms leading to extremely weak adhesion, we have developed a quantitative JKR like test, based on the formation and breakage of a capillary bridge between the investigated surface and a liquid bath. Then, capillary forces are opposed to adhesion forces, and noticeable deformations can be achieved before the rupture of contact between the liquid and the surface. Changing the distance between the liquid bath and the surface, one can tune and monitor the deformation of the capillary bridge, providing an efficient way of differentiating otherwise quite apparently similar surfaces. This test is somewhat similar in spirit to the "capillary bridge" rheometer[2,3], in which the deformation of a capillary bridge formed between two cylindrical flat punches is used to trace back the strain while the two surfaces are pulled apart at a chosen velocity. The pull force and the shape of the capillary bridge are monitored in order to extract stress – strain curves. In the capillary bridge rheometer however, the capillary bridge is deformed by changing the distance between the two surfaces at fixed contact area on each flat punch. In the present JKR like test, the contact area between the investigated surface and the liquid is let adjust when the surface is progressively pulled off the liquid bath. The triple line which delineates the edge of this contact sweeps progressively the whole surface under investigation, similarly to what happens in a conventional JKR test upon unloading. Adhesive interactions (or wetting forces) tend to maintain the liquid in contact with the surface, while capillary forces all around the capillary bridge tend to minimize the liquid –air surface and thus to decrease the contact area between the liquid and the surface. The exact shape of the capillary bridge thus results from a balance between wetting forces at the liquid – surface interface and capillary forces on the bridge, which can be smoothly changed by changing the distance between the liquid bath and the surface. The dynamics of a macroscopic capillary bridge has been used to study the nucleation and growth of a liquid meniscus but on a place surface[4]. The shape of a capillary bridge formed when a drop of liquid is inserted between two spherical objects or between a sphere and a plane has been carried out by several papers in order to obtain the adhesion force in wet systems such as granular material[5-7] or in nanoscience[8-10] since the AFM tip can often be approximated by a sphere. In the test we describe in the present paper, a very large sphere (much larger than a usual grain in a granular heap) is used and put into contact with infinite reservoir of liquid so that a behavior independent of the volume of liquid is attained. When developing this test, we have been deeply inspired by the ensemble of the work performed by Pierre-Gilles de Gennes on wetting and dewetting[11-14], and we wish, through the description of the potentialities of the capillary bridge test, to constitute a modest tribute to Pierre-Gilles' seminal scientific work.

In the present paper, we describe the principle of this capillary bridge test, and then exemplify its potentialities by investigating the formation and the rupture of capillary bridges between various fluorinated coatings and several Newtonian liquids: deionised water and silicon oils having various viscosities but almost the same surface tensions (much smaller than that of water), so that both the effect of the contact angle of the liquid on the surface and the viscosity of the liquid can be pointed out. Several parameters of the capillary bridge have revealed to be highly sensitive to small differences in the molecular organization of the coatings, especially when the pull off velocity was increased. The paper is thus organized as follow: the materials, the principle of the test and the experimental setup are presented in a first part. Then, the behavior observed in quasi static pull or push regimes is described and discussed for both water and silicon oils. In a third part, two dynamic aspects are presented and discussed: the final rupture of the capillary bridge at large pull distances, and the appearance of a thin liquid film, called the "pancake" at high enough pull off velocities. We show that these dynamic features are both highly sensitive to the coating, and can be used to characterize very weakly adhesive surfaces.

**EXPERIMENTAL SECTION**

**Materials:**

Three liquids have been used: deionised water (Millipore Milli-Q, resistivity $10^{18}$ $\Omega$-cm, surface tension $\gamma$ = 72.5 mN/m, viscosity $\eta$ = 0.997 mPa.s, density $\rho$ = 0.997 g.cm$^{-3}$), and two commercial tri methyl terminated silicon oils (Rhodia Silicone, 47V 10, surface tension $\gamma$ = 20.5 mN/m, viscosity $\eta$ = 9.4 mPa.s, density $\rho$ = 0.94 g.cm$^{-3}$ and 47V1000, surface tension $\gamma$ = 21.1 mN/m, viscosity $\eta$ = 970 mPa.s, density $\rho$ = 0.97 g.cm$^{-3}$) with a viscosity of respectively 10 and 1000 times that of water. All the above data are given at 25 °C.

The surfaces were spherical watch glasses with a radius of curvature $R$ = 10 cm, covered by three different perfluorinated coatings. All three coatings (labelled A, B and C), obtained by grafting functionalized perfluoroalkane chains by vapor deposition, have a thickness in the range 3 to 7 nm (qualitatively controlled with a quartz microbalance during the deposition process) and a small roughness of 1.5 nm, as measured through AFM in contact mode[15]. The three coatings differ essentially in the exact chemistry of the grafted polymer molecules, and cannot be detailed more accurately for confidentiality reasons. The advancing contact angles (measured by conventional contact angle measurement method) are very close and are reported in table 1, along with the contact angle hysteresis.

| System | $\theta_a$ | $\theta_r$ | $\Delta\theta$ |
|---|---|---|---|
| Coating A – water | 118.5 ± 0.4 | 88 ± 3 | 30 |
| Coating B – water | 109 ± 1.1 | 91.5 ± 1.5 | 18 |
| Coating C – water | 117 ± 1.4 | 113 ± 1.5 | 4 |
| Coating A–oil 47V10 | 53.5 ± 0.5 | 40.9 ± 1.5 | 12.6 |
| Coating B – oil 47V10 | 45.5± 1.3 | 45.5± 1.3 | 0 |
| Coating C – oil 47V10 | 49.5± 1.2 | 49.5± 1.2 | 0 |
| Coating A–oil 47V1000 | 61± 1.1 | 61± 1.1 | 0 |
| Coating B–oil 47V1000 | 53±5 | 53±5 | 0 |
| Coating C–oil 47V1000 | 54.5 | ± 0.5 | 0 |

**Table 1.** Advancing and receding contact angles and hysteresis measured by the sessile drop method for the different systems.

**Principle of the JKR like capillary bridge test:**

The weakly adhesive surface is put into contact with the liquid bath and then pulled-off at chosen velocities. The successive steps of the experiment are schematically presented in fig.1. The surface is first slowly approached to the liquid surface (A). At contact (B), a capillary bridge forms rapidly (C). The surface is then pulled-up at a chosen velocity (D) until the rupture of the capillary bridge (E). After the breakage of the capillary bridge, a small liquid drop may remain attached to the surface.

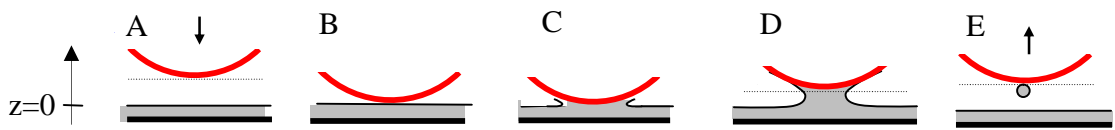

**Figure 1**. Schematic presentation of the experiment. (A): approach of the surface to the liquid bath micron by micron. As soon as the surface touches the liquid bath (B), a capillary bridge forms (C). (D): Pull-off of the surface at a chosen velocity, until the breakage of the capillary bridge. After rupture, a droplet may remain attached to the surface (E).

The evolution of the capillary bridge as a function of *z*, the distance between the surface and the liquid bath, is monitored through two synchronized video cameras providing respectively the contact area of the liquid on the surface and the profile of the capillary bridge (fig.2-I). The vertical position of the surface and the pull-off velocity are controlled through a stepper motor, with a resolution of 1 µm without hysteresis. The range of velocities is $1 \leq v \leq 500$ µm/s. Image analyses gives direct access to the contact area and the contact angle of the liquid on the surface (fig.2-II).

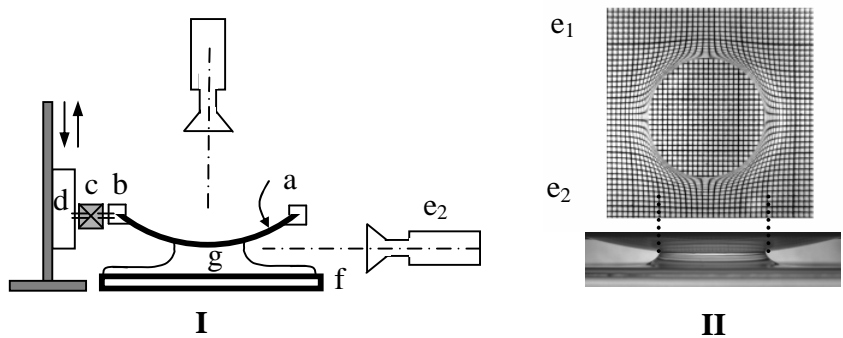

**Figure 2**. 2-I: Schematics presentation of the experimental setup. (a) transparent spherical anti-adhesive surface; (b) ring frame maintaining the position of the surface; (c) anti-vibration device in order to eliminate motor vibrations; (d) motorized vertical translation; ($e_1$) and ($e_2$) video cameras; (f) liquid bath container; (g) capillary bridge. 2-II: side view of the capillary bridge with the video $e_2$, and top view with the video $e_1$. A grid (millimetre resolution) located at the bottom of the transparent bath container is used to easily locate the edge of the contact area and quantify the contact area.

**Typical experimental cycle:**

During a pull-off motion, the contact area first remains constant, and then is observed to decrease as the distance *z* between the liquid bath and the surface is increased, in a linear manner. The capillary bridge remains stable for a large range of *z* values and of contact areas. Before breakage of the bridge, one can stop the pull motion, and reverse the motor, pushing now the surface towards the liquid bath. The contact area first remains constant when decreasing *z*, and then increases linearly with *z*.

Different cycles of push and pull can be performed, and lead to reproducible contact area versus $z$ curves, indicating a behavior independent of the history of formation of the capillary bridge. Such a typical contact area versus $z$ curve is shown in fig.3, for a perfluorinated coating in contact with deionised water. The full reproducibility is attained after the first push-pull cycle. Push-pull cycles produce two 'plateau' regimes, characterized by a constant contact area while the surface-bath distance is increased (pull) or decreased (push). All along these "plateau", the shape of the bridge evolves at fixed contact area. This is a direct signature of the contact angle hysteresis of the liquid on the surface: when the surface is pulled off the liquid, the contact line is pinned and only starts to move when the contact angle reaches the receding contact angle, while on pushing the surface towards the liquid, the contact line starts moving when the contact angle has increased and attained the value of the advancing contact angle.

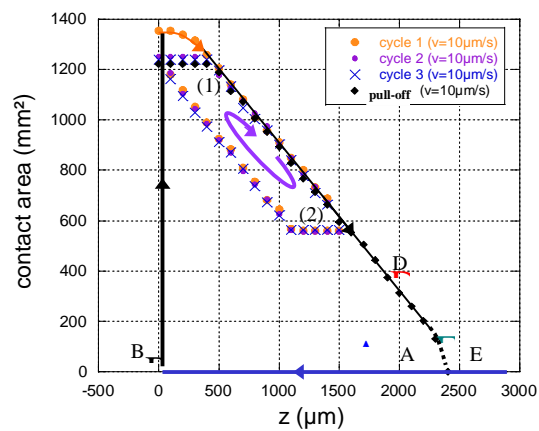

**Figure 3.** Contact area as a function of the height of the capillary bridge. The capillary bridge of deionised water is formed on an anti-adherent surface covered by a perfluorinated coating. The letters on the curves correspond to the different steps described in fig. 1.

More precisely, at the slow speed of 10µm/s (figure 3), for the particular surface used, the contact area remains constant while $z$ is increased (or decreased) by 400 µm. All along plateau labelled

(1) in fig. 3 (pull case) the contact angle decreases progressively by 10. On the contrary, the contact angle increases in the push case (plateau labelled (2) in figure 3).

**RESULTS AND DISCUSSION**

**Low velocity regime:**

The whole behavior described above is observed at low pull off velocities. We have examined how it was depending on the nature of the liquid and on the properties of the surface. Fig. 4 shows the experimental results obtained with de-ionised water and three different coatings A, B and C, all highly hydrophobic (see table 1 for the values of the contact angle with water). Fig.5 shows the experimental results obtained with deionised water or silicon oils in contact with the same coating A.

As shown in figure 4, where the contact area, in case of water, is reported as a function of $z$, the length of the "plateaux" is highly depending on the coating, clearly indicating neat differences in the contact angle hysteresis, while the slope of the contact area versus $z$ curves appears unaffected by the properties of the surface. Only the absolute position of this straight line depends on the wetting properties of the surface, the less wettable coating leading to the smallest contact area, as could have been expected. As measured during pull-off, the receding contact angles are 88±3°, 97±3° and 108±3° for respectively coating A, B and C, very close to results obtained by the sessile drop method (tab. 1).

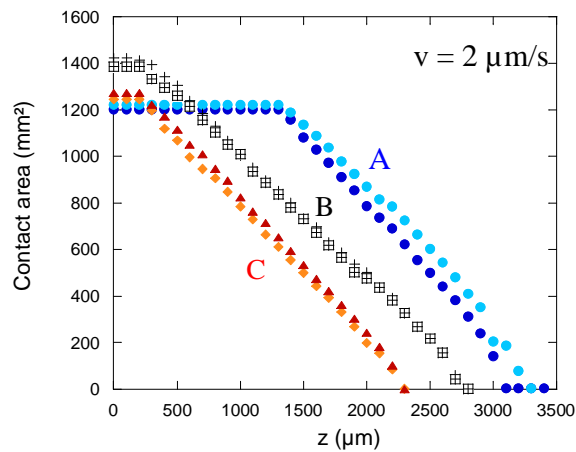

**Figure 4.** Contact area as a function of the height of the capillary bridge for deionised water and two silicon oils at low speed for three different liquids.

In order to check if the characteristics of the contact area versus distance curves were depending on the nature of the liquid, and to discriminate between wetting and hydrodynamic properties, similar experiments have been conducted with two silicon oils having quite larger viscosities than water (10 and 1000 larger), and much smaller contact angle on the fluorinated surfaces (see table 1). The behavior at low pull off velocities appears quite similar to that observed with water, as shown in figure 5 were the behaviors for water and the two silicon oils on coating A are compared. The behavior of the capillary bridge observed during pull off clearly does not depend on the viscosity of the liquid, indicating a quasi-static behavior.

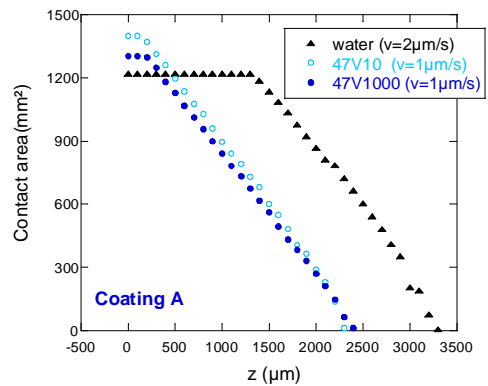

**Figure 5.** Contact area as function of the height of the capillary bridge for coating A in contact with different liquids. The coating is pulled off with a quasi-static velocity: 1-2 µm/s.

This can be further exemplified in figure 6, where data taken with coatings A and C in contact with the two silicon oils are reported for various pull off velocities. For each coating, the behavior does not depend on the liquid viscosity nor on the pull off velocity (in the range investigated here), but the absolute position of each contact area versus distance curve appears characteristic of the coating.

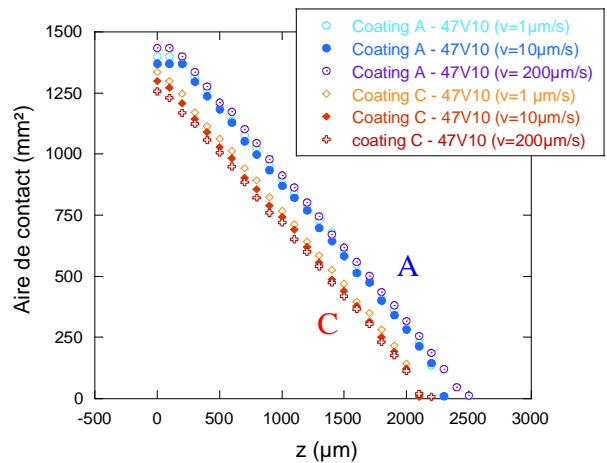

**Figure 6.** Contact area as function of the height of the capillary bridge for coating A and C, and the two silicon oils, for pull-off velocities in the range 1 to 200µm/s.

Again, it appears clear that the capillary bridge test allows an easy discrimination between coatings, the less hydrophobic coating A leading to higher contact areas than coating C at given distance between the liquid bath surface and the test surface. With the silicon oils however, the "plateaus" are less visible, indicating a smaller contact angle hysteresis than with water. The length of the plateau thus appears not to be the best stable parameter to be used when trying to discriminate between coatings, as it appears very sensitive to the nature of the test liquid.

One can understand the origin of the simple linear dependence between the wetted area on the glass and the distance between the surface and the liquid bath, $z$, and why the slope of this linear relation appears to be independent of the nature of the liquid. Let us use a simplified scaling argument: considering the capillary bridge at its formation (no large pulling), the geometry is as sketched in figure 7.

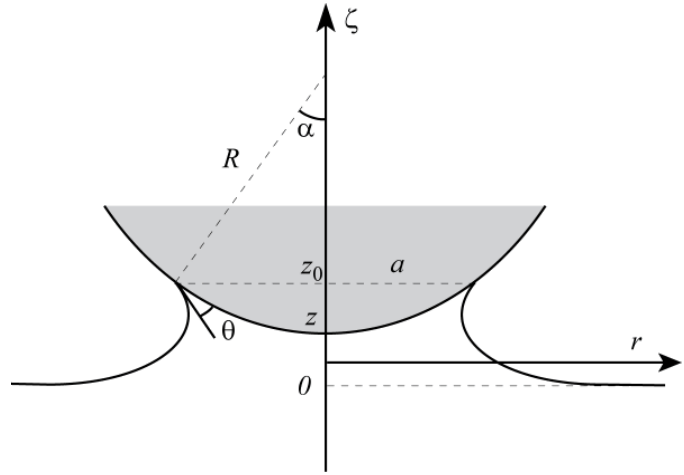

**Figure 7.** Geometrical parameters of the capillary bridge.

The radius of the glass is $R$, the radius of the wetted area is $a$, and when the glass touches the liquid bath, the liquid climbs a height $z_0$ above the rest level of the liquid bath, in order to touch the surface with the contact angle $\theta$. $z_0$ is comparable to the capillary length[16] of the liquid $\kappa^{-1} = (\gamma/\rho g)^{1/2}$ with $\gamma$ the surface tension of the liquid, $\rho$ its density and $g$ the acceleration of gravity) of the order of one millimeter, much smaller than $R$, the radius of the glass (10 cm). This height as observed in the experiments is almost constant. When the glass is pull up at a distance $z$ from the liquid bath, $a$, $z$ and $R$ remain related through the geometrical relation: $R^2 = a^2 + (R-(z_0-z))^2$, which gives, with $z_0$ and $z$ much smaller than $R$, $a^2 \sim 2R(z_0-z)$. The linear relation between the wetted area and the distance from the liquid bath results from a pure geometrical condition. The slope is fixed by the radius of the spherical glass, independently of the liquid and of the nature of the surface.

For a more precise calculation, a full analysis of the shape and of the stability of the capillary bridge can be performed numerically. The profile $r(\zeta)$ of the capillary bridge is a solution of the classic Laplace equation $1/R_1 + 1/R_2 = \Delta P/\gamma$, where $\Delta P$ is the pressure drop at the interface, and $R_1$ and $R_2$ the radii of curvature of the liquid-air interface. Using the cylindrical coordinates specified in fig.8, this equation can be rewritten as:

$$\frac{\zeta}{\kappa^{-2}} = \frac{\ddot{r}}{(1+\dot{r}^2)^{3/2}} - \frac{1}{r(1+\dot{r}^2)^{1/2}}$$

with $\ddot{r} = d^2r/d\zeta^2$ and $\dot{r} = dr/d\zeta$ This differential equation can be solved numerically, with the boundary conditions fixed by the geometry of the system: far from the spherical glass, the surface of the liquid bath is horizontal, and $\dot{r}(0) \to \infty$ the surface of the glass, the slope is defined by, $\dot{r}(z_0) = \tan^{-1}(\alpha + \theta)$ with $\theta$ the contact angle of the liquid on the surface of the glass and $\alpha = \sin^{-1}(r/R)$.

Mathematica software was used to solve this non linear differential equation. We started at a given point on the glass surface, with the contact angle θ, and use a shooting technique to find a solution which respects the boundary condition at height $z$, $\dot{r}(0) \to \infty$. For a given z, the solution is not unique, and all solutions differ in the value of the radius of the contact area, *a*, (or of the depth δ=$z_0$ – z by which the spherical glass is immersed into the liquid). To choose among the possible solutions, we have numerically estimated the energy difference Δε with and without the capillary bridge as a function of δ for each value of *z*. The results, for characteristic values of the surface tension of the liquid and of the contact angle on the surface representative of water on a perfluorinated substrate are reported in figure 8, for a curved glass, with a radius of curvature *R* = 10 cm. The curves in fig. 8(a) clearly show the role of the curvature of the glass to stabilize the capillary bridge. Indeed, on a flat surface a capillary bridge can be formed, but is not stable in the absence of contact angle hysteresis. At small *z*, gravity is negligible in front of capillarity. The component of the Laplace pressure difference between the inside and the outside of the capillary bridge associated with the radius of the bridge tends to expand it, while the other component, of opposite sign, is independent of this radius. On the curved surface, when the same component associated with the radius of the bridge tends to expand the size of the bridge, one has to pay a gravity penalty for large enough a, and this results in a stabilization of the capillary bridge at a given radius. Indeed, the numerically estimated energy versus radius of contact curves present a minimum for small enough values of *z*. The solution corresponding to that energy minimum is the stable capillary bridge for the corresponding *z* value.

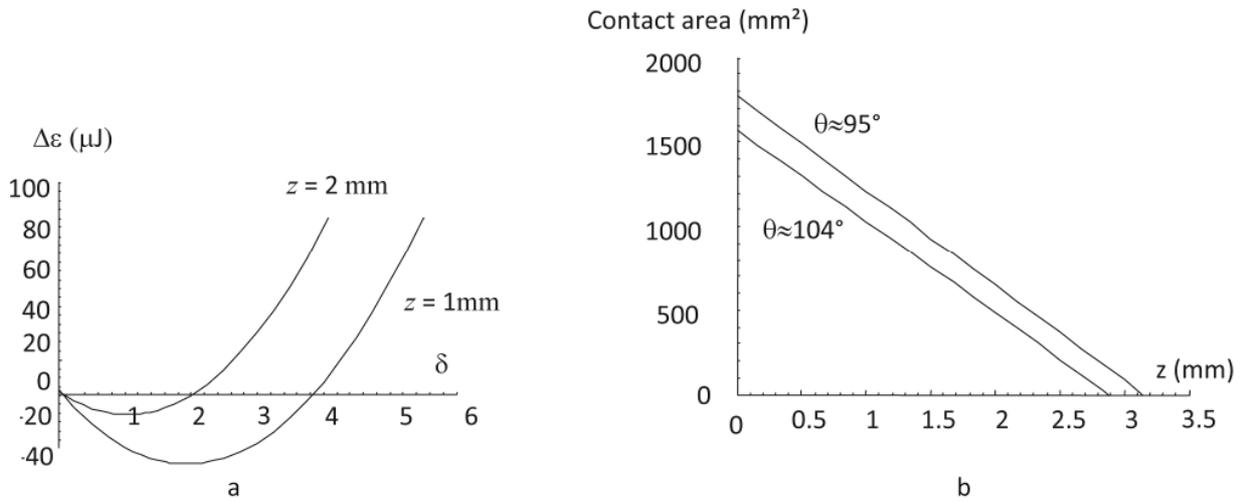

**Figure 8.** a and b: Typical example of the variation of energy with and without the capillary bridge as a function of the distance of immersion of the lens into the bath, δ, for a surface with a radius of curvature of $R = 10$ cm. The contact angle on the surface is fixed at $\theta = 85°$. (b): Contact area as a function of the height of the capillary bridge, $z$, for the optimal numerical solution of the Laplace equation.

In figure 8, the area of contact between the surface and the capillary bridge is reported as a function of $z$, the distance between the surface and the liquid bath, for the optimal numerical solutions of the Laplace equation. The results of these simulations are fully consistent with the experiment: the contact area decreases linearly with the height of the capillary bridge, and the absolute position of the straight line depends on the hydrophobic properties of the lens, with a smallest contact area for more hydrophobic surfaces. No contact angle hysteresis has been included in the above simulations. Further work is presently in progress to do so.

**Dynamic behavior: final rupture of the capillary bridge:**

The stability of the capillary bridge cannot be maintained at large $z$, and when pulling at chosen velocity, the central part of the bridge thins down and finally breaks, as shown in figure 9 for water on the three different coatings. Even if a low pull off velocity is used, the instability of the final capillary bridge develops very rapidly and one no longer deals with quasi-static phenomena. A high speed camera needs to be used to capture the final stages of the bridge.

In figure 10, one can immediately see that the bridge breaks in a very different manner depending on the coating. After the breakage of the capillary bridge, a droplet remains attached to both coatings A and B, while there is no liquid left on coating C, for which the instability leading to the breakage of the thin final capillary bridge forms right at the surface, while it forms in the middle of the bridge or close to the liquid bath in the two other cases. The total duration of these final stages, going from similar contact areas to breakage is also quite dependent on the coating, and the time needed to go to breakage follows the order of decreasing hysteresis in contact angles.

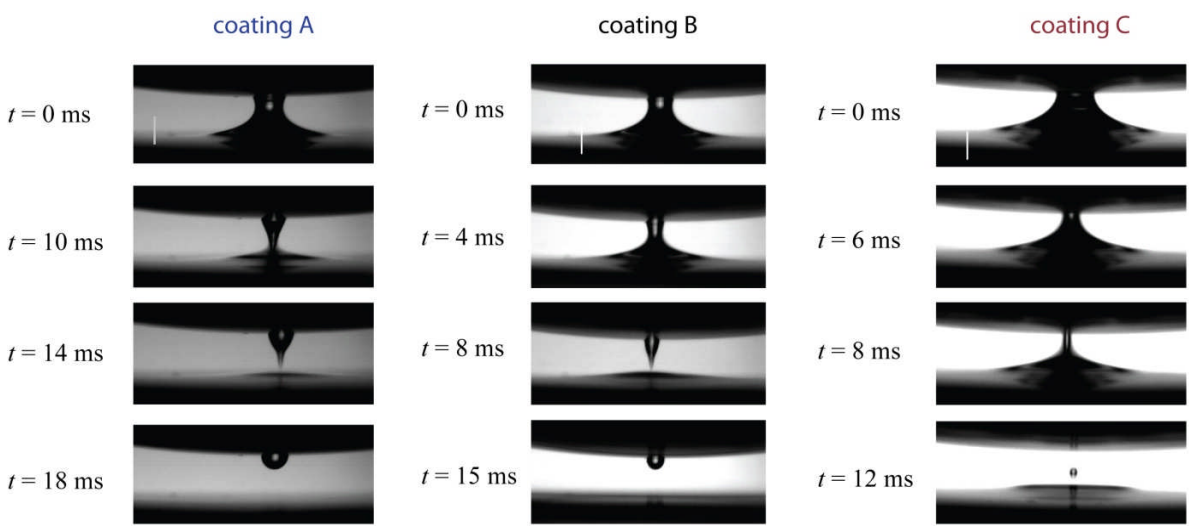

**Figure 9.** Final pictures of evolution until breakage of a capillary bridge of deionised water for the three different perfluorinated coatings analyzed in figure 5. The volume of the drop remaining on the surface depends on the properties of surface. The vertical scale bar is 2 mm long.

The volume of the drop remaining trapped on the surface after breakage, which may be a quite important quantity for applications of the coatings as anti-adhesive or easy dryable surfaces, depends both on the coating and on the pull-off velocity as shown in fig.10.

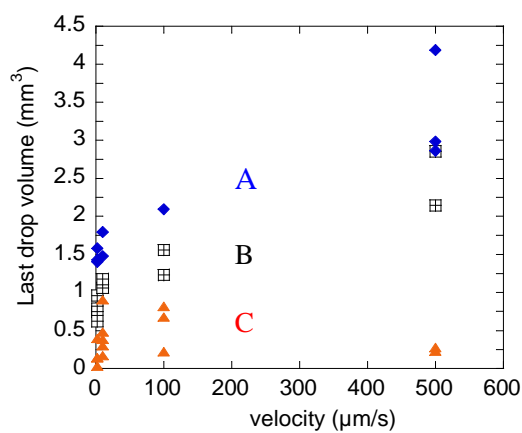

**Figure 10.** Volume of the remaining drop of de-ionized water as a function of the pull-off velocity for the three perfluorinated coatings A, B and C analyzed in quasi-static conditions in figure 5.

Although the three coatings have a close chemistry, they appear to have clearly distinct wetting properties, as revealed by the capillary bridge test both at low velocities, and at rupture. For each coating, the volume of the drop increases when increasing the pull off velocity. The coating having the highest contact angle hysteresis and the smallest receding contact angle leads to the largest volume of the remaining drop. This corresponds to the more inhomogeneous and the less hydrophobic coating.

The volume of the drop remaining on the surface appears to strongly depend on the contact angle hysteresis, and especially on the value of the receding contact angle, more than on the absolute value of the mean contact angle.

Similarly, the viscosity of the liquid deeply affects the total volume of liquid remaining on the surface at the breakage of the capillary bridge, as shown in figure 11, for the two silicon oils in contact with coatings A, B and C and a low withdrawal velocity of 1-2 µm/s..

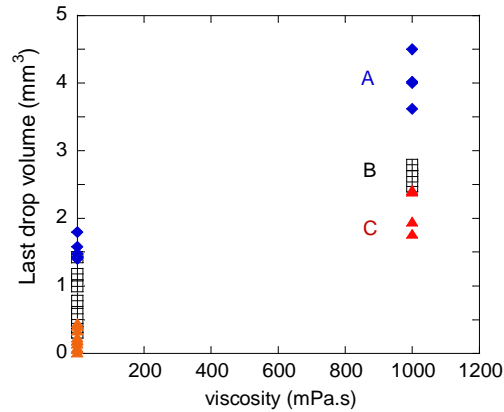

**Figure 11.** Volume of the drop remaining attached to the surface after breakage of the bridge for the two silicon oils in contact with coatings A, B and C. The pull off velocities were 1 and 2 µm/s (quasi-static pull off)

**Dynamic behavior: pull off at high velocities:**

One can also depart from the quasi-static behavior by increasing the pull off velocity. The full evolution of the contact area versus distance is then affected by the pull off velocity and the viscosity of the liquid. In figure 12, the aspect of the capillary bridge when the pull off velocity is increased (silicon oil 47V1000, pull off velocity of 500 µm/s) is reported for several pull off distances. The capillary bridge first appears similar to what is observed at low velocities (or with water or a silicon oil of low viscosity), but after a given pull off distance (1700 µm for the particular conditions of figure 12), a thin film develops, retained by the surface, while the central part of the capillary bridge goes on decreasing in size almost as if no thin film was present. We have called this thin film the pancake. Its thickness is not constant, and a ridge appears on its external edge close to the triple line.

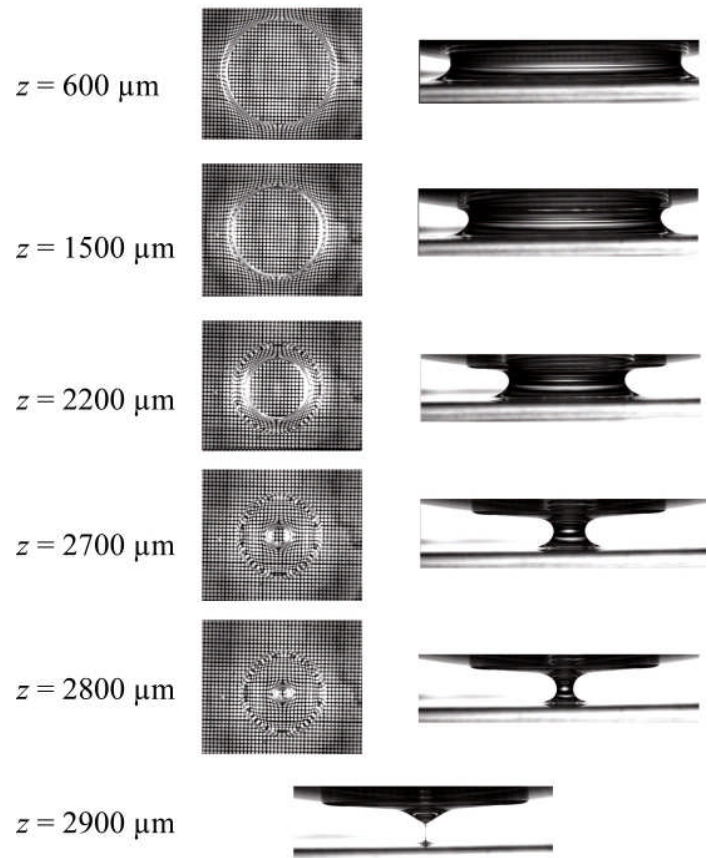

**Figure 12:** Aspect of the capillary bridge as a function of the distance between the surface and the liquid bath, when the pull off velocity is increased. Coating C is pulled off a bath of silicon oil 47V1000 at the velocity $v = 500$ µm/s. The pictures are taken by the two synchronized cameras, to give top and profiles views. The double structure of the capillary bridge, pancake and central capillary bridge is well visible on both views

In a way similar to what has been done for quasi-static conditions, one can monitor the evolution of the contact area with the distance between the surface and the liquid bath, $z$, but now two contact areas can be defined: that of the central capillary bridge and that of the pancake when it develops ahead the central capillary bridge. Such evolutions are reported in figure 13, for coating C in contact with the silicon oil 47V1000, and several pull off velocities going from 1 µm/s to 500 µm/s. Filled symbols correspond the contact area of the pancake, and the open symbols to that of the central capillary bridge.

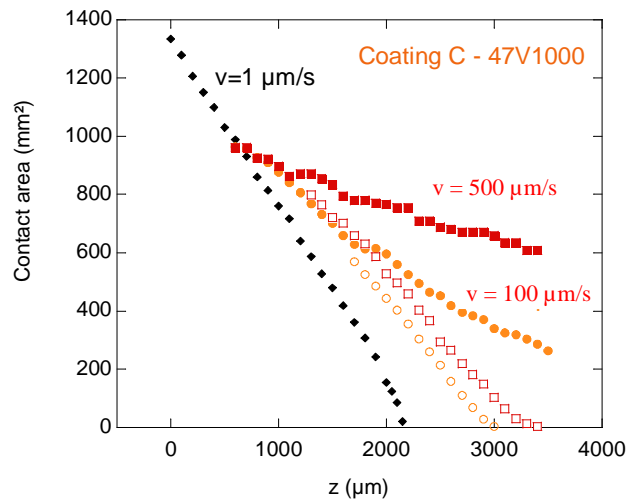

**Figure 13.** Pull off at various velocities (1, 100, et 500 µm/s) for coating C in contact with the oil 47V1000.

The quasi-static behavior is observed at the low velocity of 1 µm/s, while for much larger velocities, departures from quasi-static behavior are visible both in the appearance of the pancake and in the evolution of the central capillary bridge which now depends on the pull off velocity. The appearance of the pancake appears very similar to a Landau-Levich transition[17-20] when a solid plate is pulled off a liquid bath, a flat wetting film remains deposited on the surface above a given critical pull off velocity, connected to the liquid bath by a transition zone which profile has been described in details[20]. The retraction of the pancake should be very similar to a hydrodynamic dewetting process[21]. Further work is presently under progress to fully establish this fact.

As can be seen in figure 14, the pancake is particularly sensitive to small heterogeneities of the surface. It deforms when sweeping across these small defects of the surface, while the central part of the bridge is not affected, as it corresponds to a capillary bridge connected not to the surface but to the pancake film.

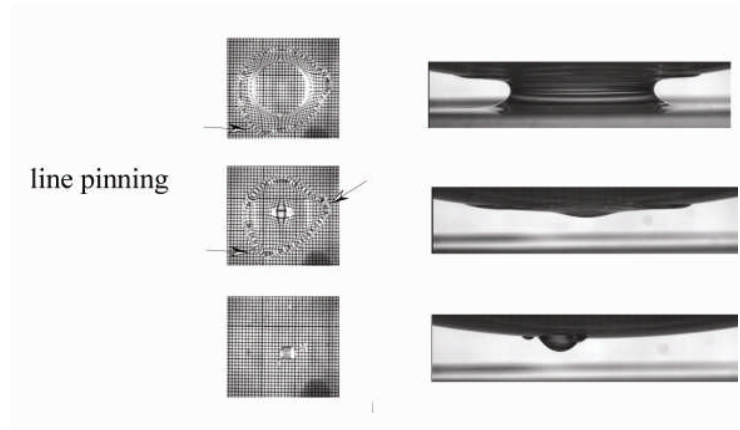

**Figure 14.** Example of pinning of the edge of the pancake by very small heterogeneities of the surface. This pinning would not have been visible in the quasi-static pull off regime.

**CONCLUSION**

We have presented a new adhesion test, the "capillary bridge test", specifically designed to analyze very weakly adhesive surfaces. In this test, deeply inspired from the now widely used JKR test, the deformations of a liquid surface are used to trace back the adhesive properties of the investigated surface. The surface to be investigated has a spherical shape, and is put into contact with a liquid bath, so that a capillary bridge forms. Pulling off the surface form the bath allows one to progressively deform the capillary bridge, and, monitoring the evolution of the capillary bridge with pull off distance yields both qualitative and quantitative information on the adhesive properties of the surface. The originality of this test is that it opposes weak capillary forces to adhesion forces, contrary to most available adhesion tests in which elastic forces are opposed to adhesive forces. We have presented in details the experimental set-up, and exemplified its potentialities by analyzing the behaviors of three slightly different perfluorinated coatings put into contact with three different liquids (water and two silicon oils having quite different viscosities). Depending of the pull off velocity, two quite different regimes can be distinguished. At low enough pull off velocity, a quasi-static capillary bridge is formed, and the evolution of the contact area of the liquid on the surface with pull off distance is linear, with a slope which only depends on the radius of curvature of the surface. The absolute position of this straight line however depends strongly on the value of the receding contact angle of the liquid on the surface, which needs to be reached to allow for the triple line to start sweeping on the surface and thus the contact area to decrease with pull off distance. Such a quasi-static capillary bridge is stable for not too large pull off distances, due to the curved geometry of the glass. At large enough pull off distances it breaks suddenly, in a dynamic manner, and a liquid drop remains attached to the surface. The volume of this remaining drop allows one to rank surfaces in terms of adhesive strength: the larger the remaining volume, the larger the tendency of the surface to develop adhesion versus any other media. For lager pull off velocities, the capillary bridge no longer evolves in a quasi-static manner, and the viscosity of the liquid becomes an important parameter. A signature of the non quasi-static behavior is the appearance of a thin liquid film, the "pancake", remaining on the surface behind the central capillary bridge. This "pancake" is very similar to a hydrodynamic wetting film. Its appearance can be qualitatively understood: when the receding velocity of the capillary bridge on the surface due to pull off becomes more rapid than the dewetting velocity on that particular surface, the triple line cannot recede fast enough, and the pancake forms. We have shown that this "pancake" was highly sensitive to tiny differences in the properties of the surface. In particular, it reveals otherwise invisible heterogeneities of the surface treatment. When the pancake is formed, the volume of liquid remaining trapped on the surface after breakage of the capillary bridge is larger than in the quasi-static pull off regime, and depends both on the viscosity of the liquid and on the wetting properties of the surface. Again, this volume is related to the adhesive power of the surface, with a sensitivity which appears enhanced compared to the quasi-static case. From a practical point of view, the capillary bridge test thus provides an easy way of objectively comparing otherwise very similar surfaces in terms of adhesive strength.

Due to his high sensitivity and convenient geometry, the above presented experiment provides a route to try to shade light in still open questions concerning the dynamics of wetting and dewetting from a more fundamental point of view. What fixes the receding contact angle on a molecular level? How is it related to the friction of the liquid on the surface? What allows a triple line to remain pinned or to sweep, at which velocity? What is the exact connection between wetting, adhesion, and friction? What is the exact role of the dissipation in the immediate vicinity of the moving triple line in the total balance of energies of the system: does it contributes, similarly to irreversible deformations near the crack tip in the case of solid adhesion, contributes significantly to the adhesive strength? The curved geometry of the surface used in the capillary bridge test allows one to smoothly sweep a triple line on the surface, avoiding edge effects, with a stable geometry for the capillary bridge. Model surfaces, with model defects could thus be used to try to investigate in details some of the questions raised more than ten years ago by Pierre-Gilles de Gennes concerning the dissipation in the immediate vicinity of a triple line, and to try to better understand the relations between adhesion and friction.

**AKNOWLEDGEMENTS**

We are deeply indebted to Pierre-Gilles de Gennes for his constant interest and help in our work on adhesion and friction phenomena, and to G. Josse, who suggested to us the potential interest of the capillary bridge test. We want to thank E. Raphaël, A. Aradian and T. Vilmin for valuable discussions on the modeling of the present experiments. We thank H. Hervet and C. Poulard for their help in the numerical analysis. We also acknowledge P. Lacan and C. Biver, from Essilor, for financial support and for providing us with the *perfluorinated* surfaces.